\documentclass{article}
\usepackage{arxiv}
\usepackage{cite}
\usepackage{amsmath,amssymb,amsfonts}
\usepackage{bm} 
\usepackage{algorithmic}
\usepackage{graphicx}
\usepackage{subfigure}
\usepackage{textcomp}
\usepackage{multirow}
\usepackage{makecell}

\graphicspath{{figures/}}

\title{Joint Trajectory Replanning for Mars Ascent Vehicle under Propulsion System Faults: A Suboptimal Learning-Based Warm-Start Approach}

\author{
	Kun Li \\
	Department of Control Science and Engineering\\
	Harbin Institute of Technology\\
	Harbin 150001, China \\
	\texttt{lkun005@stu.hit.edu.cn} \\
	\And
	Guangtao Ran \\
	Department of Control Science and Engineering\\
	Harbin Institute of Technology\\
	Harbin 150001, China \\
	\texttt{ranguangtao@hit.edu.cn} \\
	\And
	Yanning Guo \\
	Department of Control Science and Engineering\\
	Harbin Institute of Technology\\
	Harbin 150001, China \\
	\texttt{guoyn@hit.edu.cn} \\
	\And
	Ju H. Park \\
	Department of Electrical Engineering\\
	Yeungnam University\\
	Gyeongsan 38541, South Korea \\
	\texttt{jessie@ynu.ac.kr} \\
	\And
	Yao Zhang \\
	School of Engineering\\
	University of Southampton\\
	Southampton SO16 7QF, U.K. \\
	\texttt{yao.zhang@soton.ac.uk} \\
}

\begin{document}


\maketitle

\begin{abstract}
During the Mars ascent vehicle (MAV) launch missions, when encountering a thrust drop type of propulsion system fault problem, the general trajectory replanning methods relying on step-by-step judgments may fail to make timely decisions, potentially leading to mission failure. This paper proposes a suboptimal joint trajectory replanning (SJTR) method, which formulates the joint optimization problem of target orbit and flight trajectory after a fault within a convex optimization framework. By incorporating penalty coefficients for terminal constraints, the optimization solution adheres to the orbit redecision principle, thereby avoiding complex decision-making processes and resulting in a concise and rapid solution to the replanning problem. A learning-based warm-start scheme is proposed in conjunction with the designed SJTR method. Offline, a deep neural network (DNN) is trained using a dataset generated by the SJTR method. Online, the DNN provides initial guesses for the time optimization variables based on the current fault situation, enhancing the solving efficiency and reliability of the algorithm. Numerical simulations of the MAV flight scenario under the thrust drop faults are performed, and Monte Carlo experiments and case studies across all orbit types demonstrate the effectiveness of the proposed method. 
\end{abstract}

\section{Introduction}
For future Mars sample return and manned return missions, the design of the Mars ascent vehicle (MAV) is a crucial and challenging aspect. On one hand, the Martian environment has significant uncertainties, such as the impact of Martian winds and solar storms \cite{zhang2019fixed}. On the other hand, the temperature conditions on Mars are harsh, and unexpected engine and propellant issues may arise after ignition. Recent experiences indicate that rocket faults occur frequently, even without considering adverse environmental conditions, with propulsion system faults being the most common issue \cite{chang2005space}. In the event of a fault, if the rocket lacks redundancy and contingency plans, there is a risk of mission failure.

Regardless of whether faults occur, the ultimate objective is to optimize flight trajectories to the target orbit. Convex optimization \cite{boyd2004convex}, after significant development in recent years, can be widely applied to solve many trajectory optimization problems \cite{gettatelli2023convex,benedikter2019convex}. It can find the optimal solution within a finite time, effectively addressing tasks that require high real-time performance. Currently, methods such as lossless convexification and successive convexification are used to transform non-convex and nonlinear terms into convex forms \cite{accikmecse2013lossless, szmuk2020successive}. Moreover, there is a growing number of methods to improve the performance of solving convex optimization problems combined with sequential convex programming (SCP) frameworks. Hong et al. proposed a novel model predictive convex programming for a class of nonlinear systems with state and input constraints, addressing optimal guidance problems with terrain constraints \cite{hong2019model}. A combination of modified Chebyshev–Picard iteration (MCPI) and SCP is proposed to approximate dynamic equations, eliminating state variables in finite-dimensional subproblems and improving convergence performance \cite{ma2023picard}, \cite{ma2022improved}. To ensure that linearized subproblems approximate the original problem well and prevent potential infeasibility, trust-region constraints and virtual control terms were introduced in \cite{benedikter2021convex} and validated in a Falcon 9 launch vehicle simulation scenario.

When a typical fault such as a thrust drop occurs, if the flight is still conducted according to the referenced commands and trajectory, there is a high probability that the mission will fail. Therefore, it is necessary to redecide the target orbit and optimize the flight trajectory after detecting a fault using some fault diagnosis methods \cite{chen2023transfer,chen2023explainable}. Even if the original target orbit cannot be reached, entering an appropriate rescue orbit for potential rescue operations helps minimize losses \cite{zhengyu2022autonomous}. Thus, trajectory replanning after fault essentially involves two parts: redecision of the target orbit and optimization of the flight trajectory. State-triggered indicator (STI) method correlates different rescue orbits with transition states defined by three indicators, addressing the online optimization problem of target orbit and flight trajectory after thrust drop faults \cite{song2020joint}. Miao et al. introduced an auxiliary phases method that avoids linearization of the objective function and terminal constraints, enhancing the convergence performance of trajectory replanning \cite{miao2023convex}. In recent years, deep neural networks (DNN) have made significant strides in solving spacecraft trajectory planning problems \cite{chai2019six,sanchez2018real}. A DNN-based adaptive collocation method improves online trajectory replanning efficiency by establishing mappings between offline fault situations and optimal rescue orbits and terminal control variables \cite{he2022mission}.

Ensuring feasibility in each optimization during trajectory replanning is crucial. A well-set initial guess can reduce infeasible situations and improve solution efficiency. To address complex ascent problems, a three-step continuation scheme is devised to enhance the solution success rate by using the solutions from simplified problems as initial guesses for the real problems \cite{benedikter2021convex}. Banerjee et al. introduced a novel approach of using outputs from trained models to warm-start nonlinear solvers, reducing computational time while obtaining feasible and locally optimal solutions of trajectory optimization problems \cite{banerjee2020learning}.

To the best of the authors' knowledge, very few studies are currently available on MAV trajectory replanning after faults. This is a multi-phase, highly nonlinear, free-terminal time optimization problem that is inherently difficult to solve, requiring algorithms with high real-time capability. Unlike typical ascent trajectory optimization problems where the target orbit is usually predetermined, in this case, the target orbit needs to be optimized. The search space for finding the solution is large, placing high demands on the convergence of optimization algorithms. Directly using neural network mappings to derive trajectories towards the target may encounter two main issues: firstly, the mapping relationship is complex and may not be highly accurate, potentially leading to premature fuel exhaustion; secondly, neural networks rely on training data, and deviations from expected trajectories or uncertainties in the actual model can result in significant deviations in outcomes. Additionally, we cannot predict fault scenarios in advance, and it is challenging to provide reasonable initial guesses, which hinders algorithm convergence and computational efficiency \cite{xu2022predefined,li2014missile}.

The main contributions of this article are summarized as follows:
\begin{enumerate}
	\item A concise, fast, and reliable suboptimal joint trajectory replanning (SJTR) method is proposed to solve the MAV trajectory replanning problem after faults. It eliminates the need for separate decision-making on target orbits and trajectory optimization during flight. The solution to this joint trajectory replanning problem can be obtained directly instead of using state-triggered indicators for step-by-step judgment and optimization.
	
	\item A learning-based warm-start method is designed, which provides a reasonable initial guess for the SJTR method through an offline trained neural network, avoiding infeasible situations and improving solution efficiency. Additionally, it addresses the issue of inaccurate orbit type determination that can occur in some edge-case faults of the SJTR method, thereby enhancing reliability.
\end{enumerate}

The rest of this article will be organized as follows. Section II introduces the dynamics, establishing the trajectory optimization problem and target orbit redecision principles. Section III analyzes the general trajectory replanning method, the SJTR method, and the learning-based warm-start scheme. Detailed simulation results and comparative studies are presented in Section IV. Finally, Section V concludes this article.
\section{Problem Formulation}

\subsection{Dynamics Under Thrust Drop Faults}
The whole flight process can be divided into three phases which are the ascending phase, the coasting phase, and the orbiting phase. The characteristics of each phase are as follows:
\begin{enumerate}
	\item \textit{Ascending phase:} The MAV takes off with thrust provided by the first stage engines, and this phase stops when the first stage propellant is completely consumed. Thrust drop faults occur during this phase and are mainly considered as proportional drop faults, which are modeled as:
	\begin{equation}
		\label{eq_1}
		T = \left\{
		\begin{aligned}
			&{T_0},&t < {t_{\text{fail}}} \\
			&\eta {T_0},&t \ge {t_{\text{fail}}}
		\end{aligned}
		\right.
	\end{equation}
	where $t$ is the flight time, ${t_{\text{fail}}}$ is the time of fault occurrence, $0 \le\eta< 1$ represents the percentage of remaining thrust after fault, $T$ and $T_0$ are the actual and rated thrust, respectively. 

	\item \textit{Coasting phase:} No thrust is applied throughout the phase, and the end time of this phase is flexible.

	\item \textit{Orbiting phase:} The first stage separates, and propulsion is provided by the second stage engine. This phase ends when the target orbit constraints are satisfied.
\end{enumerate}

Since aerodynamic lift is negligible compared to thrust, aerodynamic drag is considered the primary form of aerodynamic force. The definition of aerodynamic drag is as follows:
\begin{equation}
	D={{C}_{d}}S\rho {{\left\| {{{\bm{v}}}_{\text{rel}}} \right\|}^{2}}/2
\end{equation}
where ${{{\bm{v}}}_{\text{rel}}}={\bm{v}}-{\bm{\omega}} \times {\bm{r}}$ is the velocity vector relative to Mars, ${{C}_{d}}$ is the drag coefficient, $S$ is the MAV’s reference area, ${\bm{\omega}} $ is the angular velocity of the Mars, ${\bm{r}}={{[{{r}_{x}},{{r}_{y}},{{r}_{z}}]}^\top}$ and ${\bm{v}}={{[{{v}_{x}},{{v}_{y}},{{v}_{z}}]}^\top}$ denote the inertial position and the velocity vectors of the MAV, respectively. The atmosphere density is considered only within an altitude of 120 km, which is modeled as $\rho (t) = {\rho _0}\exp \left( { - h(t)/{h_0}} \right)$, where $h$ denotes the altitude from the surface of Mars, $h_0$ is the Martian density scale height and ${\rho _0}$ is the atmospheric density at sea level.

We transform the thrust model (\ref{eq_1}) into a unified form, i.e., $T = \eta T_0$, $0 \le\eta\le 1$. Additionally, the thrust magnitude of the MAV is fixed, with the only adjustable parameter being the thrust direction vector ${\bm{u}}={{[{{u}_{x}},{{u}_{y}},{{u}_{z}}]}^\top}$. Establish the three-degree-of-freedom dynamics equations in the Martian Centered Inertial (MCI) coordinate system as follows:
\begin{align}
	\label{eq_4}
	&\dot {\bm{r}} = {\bm{v}}\\ 
	\label{eq_5}
	&\dot {\bm{v}} =  - \frac{\mu }{{{{\left\| {\bm{r}} \right\|}^3}}}{\bm{r}} + \frac{\eta T_0}{m}{\bm{u}} - \frac{D}{{m\left\| {{{\bm{v}}_{\text{rel}}}} \right\|}}{{\bm{v}}_{\text{rel}}}\\ 
	\label{eq_6}
	&\dot m =  - \frac{\eta T_0}{{{I_{\text{sp}}}{g_0}}} 
\end{align}
where $\mu$ is the Martian gravitational parameter, ${{g}_{0}}$ is the standard gravity, ${{I}_{\text{sp}}}$ is the specific impulse of the engine and $m$ denotes the mass of the MAV.
\subsection{Trajectory Optimization Problem}
In this subsection, we provide the mathematical description of the basic fuel-optimal trajectory optimization problem.

The linkage constraints of each phase can be expressed as:
\begin{equation}\label{eq_7}
	{{\bm{r}}_{p,0}} = {{\bm{r}}_{p-1,f}},\ {{\bm{v}}_{p,0}} = {{\bm{v}}_{p-1,f}},\ {m_{3,0}} = {m_{2,f}} - {m_{1,\text{dry}}}
\end{equation}
where $p$ $(p=2, 3)$ is the phase number, ${m_{1,\text{dry}}}$ denotes the dry mass of first stage, $(\cdot)_{p,0}$ and $(\cdot)_{p,f}$ represent the state at the initial and final times of phase $p$, respectively. 

The initial state constraint is as follows:
\begin{equation}\label{eq_7+}
	{{\bm{x}}_{1,0}} = {{\bm{x}}_0} 
\end{equation}
where $\bm{x}={{[{\bm{r}}^\top,{{{\bm{v}}}^\top},m]}^\top}$ represents the state vector consisting of position, velocity, and mass.

The terminal constraints $\bm{\phi} ({{\bm{x}}_{3,f}})$ can be expressed by referring to the five-constraint problem in \cite{lu2003closed} as
\begin{align}
	\label{eq_8}
	&{\left\| {{{\bm{r}}_{3,f}}} \right\| - a^*} = 0\\ 
	\label{eq_9}
	&{\left\| {{{\bm{v}}_{3,f}}} \right\| - \sqrt {\mu /a^*} } = 0\\ 
	\label{eq_10}
	&{{{\bm{r}}_{3,f}}^{\top}{{\bm{v}}_{3,f}}}=0\\
	\label{eq_11}
	&{{\bm{r}}_{3,f}}^{\top}{\bf{1}}_h=0\\
	\label{eq_12}
	&{{\bm{v}}_{3,f}}^{\top}{\bf{1}}_h=0.
\end{align}

The unit direction vector of the orbital angular momentum is represented by ${\bf{1}}_{h}^{\text{MCI}}=[\sin \Omega^* \sin i^*, -\cos \Omega^* \sin i^*, \cos i^*]^{\top}$, ${\bf{1}}_h$ is the representation of ${\bf{1}}_{h}^{\text{MCI}}$ in the inertial launch plumbline system. By using (\ref{eq_8})-(\ref{eq_12}), we can constrain the semi-major axis $a^*$, eccentricity $e^*$, orbital inclination $i^*$, and longitude of ascending node $\Omega^*$ of the target orbit.

The final mass ${m_{3,f}}$ must be greater than the dry mass of the second stage ${m_{2,\text{dry}}}$ and the mass of the payload ${m_\text{{payload}}}$:
\begin{equation}\label{eq_13}
	{m_{3,f}} \ge {m_{2,\text{dry}}} + m_\text{{payload}}.
\end{equation}

The magnitude of the thrust direction vector is equal to one during the ascending and orbiting phases. This constraint is often transformed through lossless convexification \cite{liu2017survey}:
\begin{equation}\label{eq_14}
	\left\| {\bm{u}}_1 \right\| \le 1, \,\left\| {\bm{u}}_3 \right\| \le 1.
\end{equation}

The performance index is the final mass of the MAV and since the thrust magnitude is constant, the performance index can be equivalently expressed as:
\begin{equation}\label{eq_15}
	J_1 = {t_{3,f}} - {t_{2,f}}
\end{equation}
where ${t_{2,f}}$ and ${t_{3,f}}$ are the final time of the coasting and orbiting phases, respectively.

In summary, the basic optimization problem can be described as P1.

\underline{\textbf{P1:}} 

Minimize: Equation (\ref{eq_15}).

Subject to: Equations (\ref{eq_4})-(\ref{eq_14}).

\subsection{Target Orbit Redecision Principles}
Without considering faults, the target orbit is usually predetermined; however, when a thrust drop fault occurs, if the MAV still follows the original trajectory, there is a risk that the target orbit cannot be reached or result in mission failure. It is necessary not only to design an optimal flight trajectory, but also to evaluate the capability to achieve the orbit according to the fault situation and the actual state of the MAV in advance, so as to complete the redecision of the target orbit.

The propellant consumption for adjusting the orbital plane is much larger than that for adjusting the orbital altitude. Referring to the definition of the STI, in order to ensure safety, we should prioritize the orbit altitude when redeciding the target orbit, and the adjustment of the orbital plane will be considered only when the remaining capacity is greater than a certain threshold. The target and rescue orbits studied in this paper are both circular orbits, and the specific types of entry into orbit corresponding to the redecision principles can be categorized into four types:

\begin{enumerate}
	\item \textit{Original target orbit:} Small fault condition, even if a thrust drop fault occurs, the remaining propellant is able to support the MAV into the original target orbit.
	
	\item \textit{Rescue orbit type I:} Medium fault condition, cannot enter the original target orbit, but can reach the same altitude as the target orbit. While maintaining its orbital altitude and further reducing the deviation of the orbital plane to make the rescue orbit as close as possible to the target orbit.
	
	\item \textit{Rescue orbit type II:} Large fault condition, cannot enter the original target orbit or reach the same altitude as the target orbit. At this time, only the deviation of the orbit altitude from the original target orbit altitude is reduced, without considering the orbital plane deviation.
	
	\item \textit{Mission failure:} Severe fault condition, the MAV cannot reach the minimum safe orbit altitude.
	
\end{enumerate}

\section{Trajectory Replanning After Faults}
Trajectory replanning after thrust drop fault is an optimal joint optimization problem of rescue orbit and flight trajectory. However, it is tough to solve these two coupled problems directly. This is because different fault scenarios correspond to different types of rescue orbits, making it challenging to represent the problem in a unified form. This may lead to non-convergence of the optimization algorithm or low computational efficiency. Since the rescue orbit serves as the target orbit for flight trajectory optimization and is unknown, this greatly expands the search space for the optimal solution and makes it difficult to provide an idea initial guess.
\subsection{General Trajectory Replanning Method}

Drawing on the multi-step optimization framework based on STI, this subsection utilizes the general trajectory replanning method to construct the procedure that follows the principle mentioned in the previous section for solving the problem. 

First, without considering the semi-major axis, orbital inclination, and longitude of ascending node constraints at the final time, (\ref{eq_9}) is changed to:
\begin{equation}\label{eq_16}
	{\left\| {{{\bm{v}}_{3,f}}} \right\| - \sqrt {\mu /a_f} } = 0
\end{equation}
where $a_f =  \left\| {{{\bm{r}}_{3,f}}} \right\| $ is the magnitude of the semi-major axis at the final time. The optimization problem is then transformed into P2, which optimizes the highest circular orbit.

\underline{\textbf{P2:}} 

Minimize: $J_2 = -a_f$.

Subject to: Equations (\ref{eq_4})-(\ref{eq_7}), (\ref{eq_10}), (\ref{eq_13})-(\ref{eq_14}), (\ref{eq_16}).

When $a_f$ is less than the safe orbit altitude $a_{\text{safe}}$, it indicates mission failure. When $a_{\text{safe}} \le a_f \le a^*$, the optimized result is considered the highest circular rescue orbit, classified as rescue orbit type II. If $a_f>a^*$, it means that the MAV can reach the original target orbit altitude while further reducing the deviation of the orbital inclination $i_f$ and longitude of ascending node $\Omega_f$. In this case, a new problem P3 needs to be solved, where it is necessary to ensure $a_f = a^*$. P3 is used to determine the rescue orbit that minimizes the orbital plane difference at the same altitude as the original target orbit.

\underline{\textbf{P3:}} 

Minimize: $J_3 = \lambda_i |i_f-i^*| + \lambda_{\Omega} |\Omega_f-\Omega^*|$.

Subject to: Equations (\ref{eq_4})-(\ref{eq_10}), (\ref{eq_13})-(\ref{eq_14}).

In P3, $\lambda_i$ and $\lambda_{\Omega}$ are the weights. This problem can simultaneously correspond to entering the original target orbit and the rescue orbit type I. When the performance index is small enough, it implies that the semi-major axis and the orbital plane elements both meet the requirements of the target orbit. Otherwise, it belongs to the rescue orbit type I.

Now, P2 and P3 have been formulated for the scenarios corresponding to the four types of orbit. The process of solving this issue involves initially considering the worst-case scenario and then gradually ``upgrade" the mission. In contrast, a method of ``downgrade" is proposed to first consider the best-case scenario \cite{miao2023convex,miao2021successive}. Unlike the upgrade scheme, this downgrade method initially does not constrain the final mass of the MAV and its aim is to first obtain a set of solutions that satisfy all of the target orbital constraints. The steps for the downgrade scheme are as follows. We define a new problem:

\underline{\textbf{P4:}} 

Minimize: $J_4 = -m_{3,f}$.

Subject to: Equations (\ref{eq_4})-(\ref{eq_12}), (\ref{eq_14}).

By solving P4, if the obtained solution satisfies (\ref{eq_13}), it indicates that the MAV has sufficient fuel to enter the original target orbit. Conversely, it means the MAV needs to consume more fuel than available, thus necessitating entry into a rescue orbit or resulting in mission failure. If entering a rescue orbit is required, the next step is to solve P3. If the problem is feasible, the resulting trajectory is for entering the rescue orbit type I. If the problem is infeasible, then determine whether it is possible to enter the rescue orbit type II by solving P2 to obtain the highest rescue orbit. If this highest rescue orbit exceeds the altitude of the safe orbit, then this is the solution for the rescue orbit type II. Otherwise, the mission is deemed a failure.

\begin{figure*}[htbp]
	\centerline{\includegraphics[width=38pc]{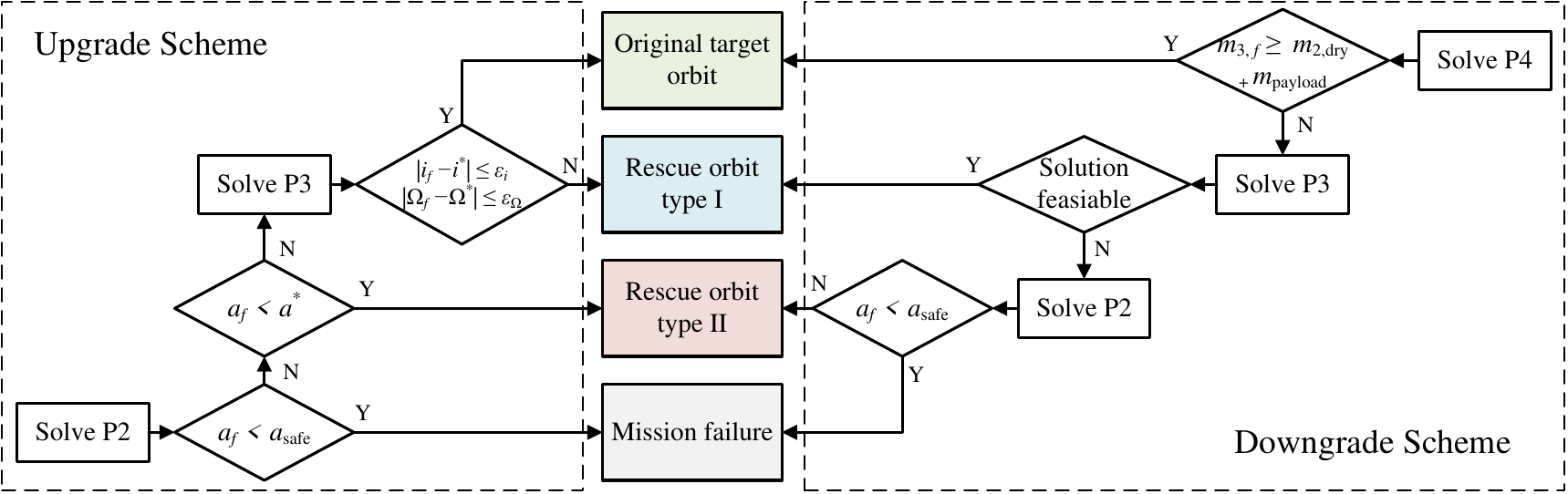}}
	\caption{General trajectory replanning method.}
	\label{fig.1}
\end{figure*}

The general trajectory replanning method is shown in Fig. \ref{fig.1}. In practical applications, the upgrade or downgrade strategies can be flexibly chosen based on the specific requirements.

\subsection{SJTR Method}
Due to the unknown rescue orbit, the search space for the optimal solution is enormous. Consequently, the general trajectory replanning method requires solving the optimization problem step-by-step, which may lead to time-consuming and infeasible solution propagation. Also, the divergence in search directions caused by conflicts between the orbital plane and shape elements, significantly reduces solution efficiency or rendering infeasible. For typical trajectory optimization problems, precision and optimality are crucial. However, for the specific problem of trajectory replanning after faults studied in this paper, convergence and computational efficiency are more important. MCPI is a method known for its good convergence and computational efficiency \cite{bai2010modified, ma2022feasible}. In this subsection, we developed the SJTR method within the MCPI framework.

For solving an initial value problem of a nonlinear differential equation, we usually use the integral form for calculation:
\begin{equation}
	{\bm{x}}(t)={\bm{x}}\left( {{t}_{0}} \right)+\int_{{{t}_{0}}}^{t}{\bm{f}}(s,{\bm{x}}(s))\text{d}s,\,t\in [{{t}_{0}},{{t}_{f}}].
\end{equation}

In contrast, MCPI offers a new way of solving the problem. The state trajectories are approximated by Picard iteration, and then continuously iterated to ensure their computational accuracy. The formula for Picard iteration is as follows:
\begin{equation}\label{eq_19}
	{{{\bm{x}}}^{k}}(t)={\bm{x}}\left( {{t}_{0}} \right)+\int_{{{t}_{0}}}^{t}{{\bm{f}}}\left( s,{{{\bm{x}}}^{k-1}}(s) \right)\text{d}s,\,k=1,2,\ldots 
\end{equation}
where ${{\bm{x}}^{k}}$ denotes the solution of $k^{th}$ iteration in the successive solution process. We perform $t=({{t}_{f}}+{{t}_{0}})/{2} +({{t}_{f}}-{{t}_{0}}) \tau /{2}$ to transform the time domain from $t \in [{{t}_{0}},{{t}_{f}}]$ to $\tau \in [-1,1]$, and obtain the new Picard iteration formula:
\begin{equation}
	{{{\bm{x}}}^{k}}(\tau )={{{\bm{x}}}_{0}}+\int_{-1}^{\tau }{\bm{g}(s,{{{\bm{x}}}^{k-1}}(s))\text{d}s},\,k=1,2,\ldots. 
\end{equation}

Use $N$ Chebyshev-Gauss-Lobatto nodes for discretization:
\begin{equation}
	{{\tau }_{j}}=-\cos (j\pi /N),\quad j=0,1,2,\ldots ,N.
\end{equation}

We can get the approximated force function by using orthogonal Chebyshev polynomials:
\begin{equation}
 	\bm g(\tau ,{{{\bm{x}}}^{k-1}}(\tau ))=\sum\limits_{i=0}^{N-1}{{\bm{F}}_{i}^{k-1}}{{T}_{i}}(\tau )  
\end{equation}
where ${{T}_{i}}(\tau )=\cos (i\arccos (\tau ))$ represents the basis function. Based on the discrete orthogonality of the Chebyshev polynomials, we can compute the coefficient vector $\bm{F}_i^{k-1}=({1}/{c_i})\sum_{j=0}^Nz_j \bm{g}(\tau_j,\bm{x}^{k-1}(\tau_j))T_i(\tau_j)$, $c_0=N;\: c_i=N/2,\:\text{for}\: i=1,2,...N;\:z_0=z_N=1/2;\:z_j=1\;\text{for}\; j=1,2,...N-1.$

The Picard iteration formulas for velocity and position are as follows:
\begin{equation}\label{eq_24}
	{{{\bm{v}}}_{p}^{k+1}}({{\tau }_{p}})\approx {{{\bm{v}}}_{p,0}}+\int_{-1}^{{{\tau }_{p}}}{\frac{({{t}_{p,f}}-{{t}_{p,0}})}{2}}{{\bm{f}}_{{\bm{v}},p}}({\bm{x}}_{p}^{k},{{{\bm{u}}}_{p}})\text{d}s
\end{equation}
\begin{equation}\label{eq_25}
	{{\bm{r}}_{p}^{k+1}}({{\tau }_{p}})= {{{\bm{r}}}_{p,0}}+\int_{-1}^{{{\tau }_{p}}}{\frac{({{t}_{p,f}}-{{t}_{p,0}})}{2}}{\bm{v}}_{p}^{k+1}\text{d}s
\end{equation}
where ${{t}_{p,0}}$ and ${{t}_{p,f}}$ are the initial and final times of phase $p$, respectively. ${{\bm{f}}_{{\bm{v}},p}}$ is the velocity parts of the function ${\bm{f}}$, ${\bm{x}}_{p}^{k}$ is the reference state trajectory of the $k^{th}$ iteration and ${{\bm{u}}_{p}}$ is the control history.

By using (\ref{eq_24}) and (\ref{eq_25}) to compute the state, we do not need to integrate the state variables. Instead, we can compute it based on the result of the previous iteration and the control quantity, and the states at each step are not coupled with each other. This changes the optimization variables from $\bm{\Xi}_0$ (${{{\bm{x}}}_{1}},{{{\bm{x}}}_{2}},{{{\bm{x}}}_{3}},{{{\bm{u}}}_{1}},{{{\bm{u}}}_{3}},{{t}_{2,f}},{{t}_{3,f}}$) to $\bm{\Xi}_1$ (${{{\bm{u}}}_{1}},{{{\bm{u}}}_{3}},{{t}_{2,f}},{{t}_{3,f}}$). It simplifies the optimal control problem, improves the convergence performance of the algorithm and expands the feasible domain.

The discrete form of (\ref{eq_24}) and (\ref{eq_25}) can be expressed as:
\begin{equation}
	{{\bm{v}}_p}[n] = {{\bm{v}}_p}[0] + {\frac{({{t}_{p,f}}-{{t}_{p,0}})}{2}}{\left[ {{{\bm{R}}_{[n + 1, \cdot ]}} \cdot {\bm{Y}} \cdot {\bm{G}}\left( {{\bm{X}}_p^k,{{\bm{U}}_p}} \right)} \right]^{\top}}
\end{equation}
\begin{equation}
	\begin{aligned}
		{{\bm{r}}_p}[n] = {{\bm{r}}_p}[0] + {\frac{({{t}_{p,f}}-{{t}_{p,0}})}{2}}\left( {{\tau _p}[n] + 1} \right){{\bm{v}}_p}[0]+ {\frac{({{t}_{p,f}}-{{t}_{p,0}})^2}{4}}{\left[ {{{\bm{R}}_{[n + 1, \cdot ]}} \cdot {\bm{Y}} \cdot {\bm{R}} \cdot {\bm{Y}} \cdot {\bm{G}}\left( {{\bm{X}}_p^k,{{\bm{U}}_p}} \right)} \right]^{\top}}
	\end{aligned}
\end{equation}
where the discrete nodes $n=1,2,...,N$, the coefficient matrices ${\bm{R}}$ and ${\bm{Y}}$ are constants when the number of discrete nodes is fixed, their computation method can be seen in \cite{junkins2013picard}. ${{{\bm{R}}}_{[n+1,\cdot ]}}$ denotes the $(n+1)^{th}$ row of ${\bm{R}}$. 

The force function matrix is given by
\begin{equation}
	{\bm{G}}( {\bm{X}}_{p}^{k},{{{\bm{U}}}_{p}})=\left[ 
		{{{{\bm{f}}}_{{\bm{v}},p}}({\bm{x}}_{p}^{k}[0],{{{\bm{u}}}_{p}}[0])},...,{{{{\bm{f}}}_{{\bm{v}},p}}({\bm{x}}_{p}^{k}[N],{{{\bm{u}}}_{p}}[N])}
	 \right]^{\top}.
\end{equation}

Since the terminal constraints are functions with respect to the state quantities, we need to convert them into functions of the control quantities and time after using the MCPI method. In order to solve the problem in the SCP framework, we need to linearize the nonconvex terminal constraints $\bm{\phi} ({{\bm{x}}_{3,f}})$ with first-order Taylor expansion, which yields:
\begin{equation}
	\begin{aligned}
		\bm{\phi}' ({\bm{x}}_{3}[N]) \approx {\bm{\phi}} ({\bm{x}}_3^k[N]) + {\nabla _{\bm{x}}} {\bm{\phi}} ({\bm{x}}_3^k[N]) \left[ \frac{\partial {{\bm{x}}_3}[N]}{\partial {{\bm{u}}_1}}({{\bm{u}}_1} - {\bm{u}}_1^k) \right. \\
		+ \left. \frac{\partial {{\bm{x}}_3}[N]}{\partial {{\bm{u}}_3}}({{\bm{u}}_3} - {\bm{u}}_3^k) + \sum_{p=2}^3 \frac{\partial {{\bm{x}}_3}[N]}{\partial t_{p,f}}(t_{p,f} - t_{p,f}^k) \right] = 0
	\end{aligned}
\end{equation}
where 
\begin{equation}
	{\nabla _{\bm{x}}}{\bm{\phi}} ({\bm{x}}_3^k[N]) = \left.\frac{\partial \bm{\phi} ({\bm{x}}_3[N])}{\partial {{\bm{x}}_3[N]}}\right|_{{\bm{x}}_3[N]={\bm{x}}_3^k[N]}.
\end{equation}

However, it is difficult to satisfy this terminal equality constraint (\ref{eq_8})-(\ref{eq_12}), and it can encounter artificially infeasible situations. To address this issue, a 5-dimensional slack variable ${{\bm{\Delta} }_{\bm{\phi} }}$ is introduced to relax the terminal constraints. 
\begin{equation}\label{eq_31}
	 \left| \bm{\phi}' ({\bm{x}}_{3}[N]) \right| \le {{\bm{\Delta} }_{\bm{\phi} }}.
\end{equation}

To limit the size of ${{\bm{\Delta} }_{\bm{\phi} }}$, the objective function of the original trajectory optimization problem becomes:
\begin{equation} \label{eq_33}
	J_5=J_1+J_{\bm{\phi }}
\end{equation}
where $J_{\bm{\phi }}=\bm{\omega}_{\bm{\phi }}^{\top}{{\bm{\Delta }}_{\bm{\phi }}}$ and $\bm{\omega}_{\bm{\phi }}$ is a positive penalty parameter. 

The application of slack variables not only relaxes the original strict equality constraints, but also reveals that the objective function $J_{\bm{\phi }}$ here is structurally similar to the objective function in P3. Therefore, the SJTR method discovers and utilizes this characteristic, adjusting penalty coefficients to preferentially optimize different orbital parameters.

Specifically, according to the target orbit redecision principle, we should set the penalty coefficients in $\bm{\omega}_{\bm{\phi }}$ corresponding to the semi-major axis and eccentricity to larger values. This ensures that the requirement for the semi-major axis of the circular orbit is prioritized when solving the problem. This way, we can achieve the joint optimization of target orbit that satisfies the redecision principle and flight trajectory directly, without the need to trigger step-by-step optimization. Since the original constraints have been relaxed and the weight of $J_1$ in the objective function is small, this essentially represents a suboptimal method. Such suboptimal methods are permissible in this paper, as it has been mentioned earlier that ensuring convergence and efficiency are more important than optimality.

Let $\bm{\Lambda} =[{{{\bm{u}}}_{1}}^{\top},{{{\bm{u}}}_{3}}^{\top},{{t}_{2,f}},{{t}_{3,f}}]^{\top}$. Trust-region constraints are imposed to ensure the feasibility of linearization:
\begin{equation}\label{eq_34}
	\left|\bm{\Lambda}_1-\bm{\Lambda}_1^k \right|\le \bm{\delta}
\end{equation}
where $\bm{\Lambda}_1^k$ is the solution of the previous optimization iteration and $\bm{\delta}$ defines the radius of trust-region. In order to further improve the convergence performance, we adopt the adaptive trust-region strategy mentioned in \cite{li2024adaptive}, which adjusts the size of the trust-region radius at different iteration stages.

In summary, the optimal control problem constructed using the SJTR method is as follows:

\underline{\textbf{P5:}} 

Minimize: Equation (\ref{eq_33}).

Subject to: Equations (\ref{eq_14}), (\ref{eq_31}), (\ref{eq_34}).

Using the SJTR method may lead to the following scenarios and corresponding types of orbits:

\begin{enumerate}
	\item \textit{Original target orbit:} The slack variable  ${{\bm{\Delta} }_{\bm{\phi} }}$ converge to a value that can be considered equal to zero, meaning the actual terminal conditions deviate negligibly from the target orbit.
	
	\item \textit{Rescue orbit type I:} Due to the large penalty coefficients associated with the semi-major axis and eccentricity errors, the corresponding slack variables can converge to negligible values. However, the errors in the orbital inclination and longitude of ascending node are non-negligible and are considered primary performance indicators in the optimization problem.
	
	\item \textit{Rescue orbit type II:} All slack variables ultimately converges to non-negligible values, but the minimum safe altitude requirement can be satisfied, and the optimization is still dominated by adjusting the altitude.
	
	\item \textit{Mission failure:} The obtained slack variables are so large that the maximum altitude is less than $a_\text{safe}$, or the optimization results are infeasible.
\end{enumerate}

\subsection{Learning-based Warm-start}
\begin{figure}[h]
	\centerline{\includegraphics[width=20pc]{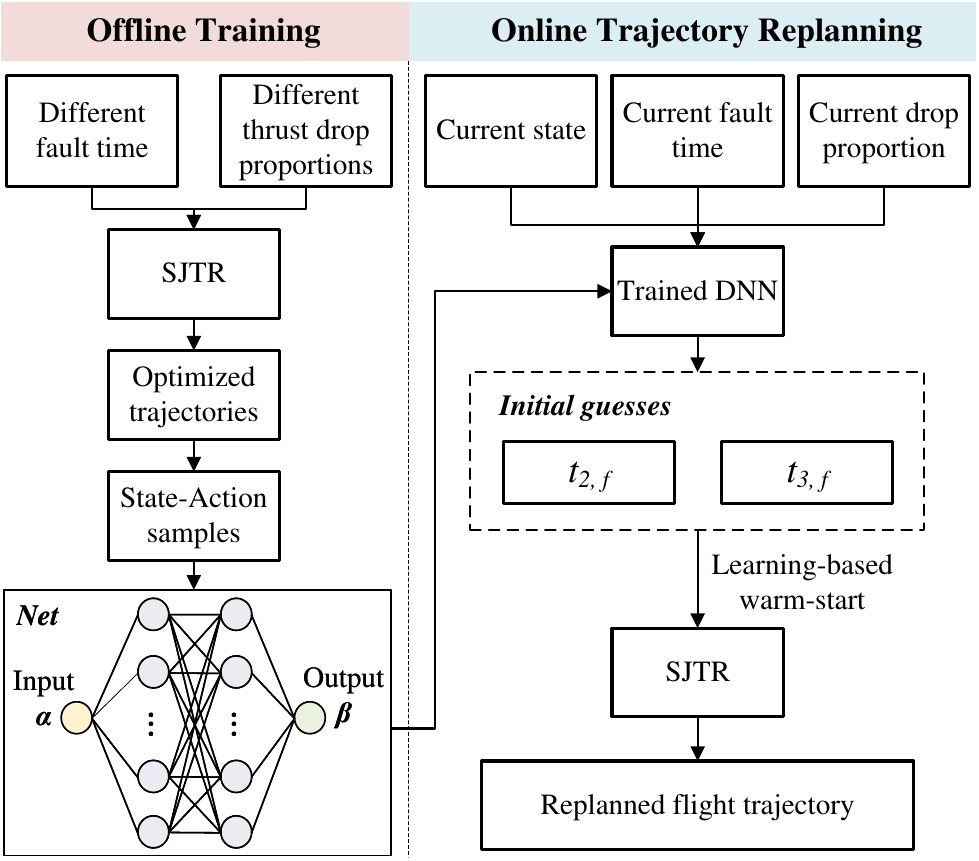}}
	\caption{Learning-based warm-start SJTR method.}
	\label{fig.2}
\end{figure}

Although SJTR has exhibited good convergence performance, an accurate initial guess can effectively improve the reliability and efficiency of the solution, and prevent infeasible situations in extreme cases. Accurate terminal time guesses are crucial for optimization, as they affect the difficulty of solving optimization problems. Therefore, we designed the learning-based warm-start method for SJTR to avoid using random initial time guesses. The overall scheme is shown in Fig. \ref{fig.2}.

This learning-based warm-start method can be divided into offline and online components. In the offline part, the dataset is divided into training data and validation data. The training data is used to train the neural network model, and the validation data is used to assess the effectiveness of the trained model. By training on a dataset of trajectory optimization solutions under various fault scenarios, we establish a mapping between the current state, fault conditions, and the final time of the coasting and orbiting phases. In the online part, the pre-trained mapping allows for rapid determination of flight time for these two phases, providing a more accurate initial guess for subsequent trajectory replanning. The DNN’s input $\bm{\alpha}$ and output $\bm{\beta}$ are as follows:
\begin{equation}
	\bm{\alpha} = {[{\bm{x}}_{\text{fail}}^{\top}, {t_{\text{fail}}}, \eta]}^{\top}
\end{equation}
\begin{equation}
	\bm{\beta} = [{t_{2,f}}, {t_{3,f}}]^{\top}
\end{equation}
where ${\bm{x}}_{\text{fail}}$ is the state vector of the MAV at the time of fault.

In order to train effectively and achieve rapid convergence, the Z-Score normalization method is employed to standardize both input and output \cite{shi2020deep}. The process is as follows:
\begin{equation}
	z = \frac{\zeta - \bar \zeta}{\sigma}
\end{equation}
where $\bar \zeta$ and $\sigma$ denote the mean and standard deviation of the training dataset $\zeta$, respectively.

The designed DNN is a fully connected feedforward neural network with an input layer, multiple hidden layers, and an output layer. The sigmoid function is used as the activation function and the mean absolute error is used to evaluate the deviation between predicted optimization values and outputs. Moreover, the Adam algorithm is employed to minimize the loss function.

In this way, when a fault occurs, we can predict the flight time of the coasting and orbiting phases for the current state based on the time of the fault and the percentage of thrust remaining. The computational time for this process is often negligible. We do not require highly accurate predictions, as long as they are in the vicinity of the optimal value, they can provide a good warm-start for the SJTR algorithm.

\section{Simulations and Analysis}
In this section, the simulation results of applying the learning-based warm-start SJTR method to the trajectory replanning problem for a Mars ascent vehicle after a thrust drop fault are presented. The assignment of MAV parameters is displayed in Table \ref{tab_1}, where ${m_{(\cdot),\text{dry}}}$, ${m_{(\cdot),\text{prop}}}$, ${{I}_{(\cdot),\text{sp}}}$, $T_{(\cdot)}$ and ${{t}_{(\cdot),\text{std}}}$ correspond to the dry mass, the propulsion mass, the specific impulse, the thrust magnitude and the burning time without faults, respectively. The payload mass of the MAV $m_\text{{payload}}= 5$ kg, and the total mass $m_0$ at takeoff is 350 kg.
\begin{table}[htbp]
	\centering
	\caption{MAV parameters}
	\renewcommand{\arraystretch}{1.2}
	\begin{tabular}{c c c c}
		\hline
		Parameters & Value & Parameters & Value \\
		\hline
		${m_{1,\text{dry}}}$, kg &  27.6 & ${m_{2,\text{dry}}}$, kg& 70.4 \\
		${m_{1,\text{prop}}}$, kg & 196 & ${m_{2,\text{prop}}}$, kg & 51 \\
		${{I}_{1,\text{sp}}}$, s & 293 & ${{I}_{2,\text{sp}}}$, s & 315 \\
		$T_1$, N & 9000 & $T_2$, N & 800 \\
		${{t}_{1,\text{std}}}$, s & 64.17 & ${{t}_{2,\text{std}}}$, s & 196.93 \\
		\hline
	\end{tabular}
	\label{tab_1}
\end{table}

The initial position of the MAV in the MCI coordinate system is ${\bm{r}_{1,0}} = {[-303.103, -3374.249, 238.074]^{\top}}$ km. We set the parameters of the target orbit as ${a^*} = (300+R_M)$ km, ${e^*} = 0$, ${i^*} = 29.5$ deg and ${\Omega^*} = 253.2$ deg, where $R_M=3396.19$ km represents the radius of Mars, and the minimum safe orbital altitude is $a_\text{safe} = 250$ km.
\subsection{Analysis of the Results for Learning-based Warm-start SJTR Method}
The number of Chebyshev nodes is $N=100$, and the initial radius of trust-region constraint is $\bm{\delta}=[0.5,0.5,6,6]^{\top}$. During optimization iterations, if the results of two consecutive iterations are less than the convergence tolerance ${{{\bm{\epsilon}} }_{{\bm{x}}}} = {[1{\rm{m,\ }}1{\rm{m,\  }}1{\rm{m,\ }}0.1{\rm{m/s,\ }}0.1{\rm{m/s,\ }}0.1{\rm{m/s,\ }}0.1{\rm{kg}}]^{\top}}$, ${{{\bm{\epsilon}} }_{{{{\bm{u}}}_{1}}}} = {{{\bm{\epsilon}} }_{{{{\bm{u}}}_{3}}}} = {[{\rm{0.001,\ }}{\rm{0.001,\ }}{\rm{0.001}}]^{\top}}$ and ${\epsilon _{{t_{2,f}}}}{\rm{=\ }}{\epsilon _{{t_{3,f}}}} = 0.1{\rm{s}}$, the optimization process is considered to have successfully concluded. The penalty coefficients for terminal constraints are set to $\bm{\omega}_{\bm{\phi }}=[10^7,10^7,10^7,6\times10^5,6\times10^5]^{\top}$.

The initial trajectory guess uses linear interpolation from the starting point to the endpoint. To generate an offline dataset of DNN, we performed 100,000 sets of Monte Carlo simulations using the SJTR method, where each set of data results from random fault occurrence time ${t_{\text{fail}}}\in[0,{{t}_{1,\text{std}}}]$ and remaining thrust percentage $\eta\in[0.15,0.5]$, both of which follow a uniform distribution. The data is divided into a training set consisting of 90,000 trajectories and a test set consisting of 10,000 trajectories. The DNN has two hidden layers and the number of neurons in both layers is 64. There is a total of 50 batches in training, and the number of samples per batch was set to 128. The learning rate is set as 0.01.

\begin{figure}[htbp]
	\centering
	\subfigure[Errors of ${t_{2,f}}$]{
		\label{fig.tf2error}
		\includegraphics[width=0.35\textwidth]{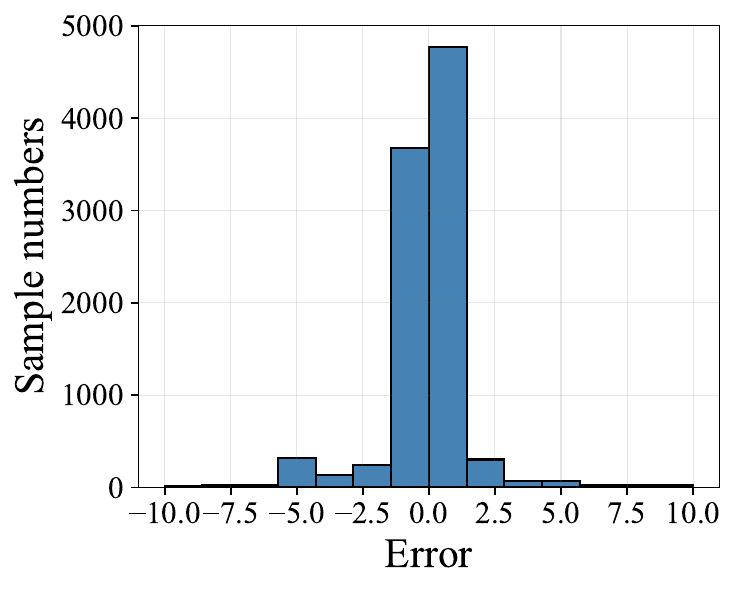}}
	\subfigure[Errors of ${t_{3,f}}$]{
		\label{fig.tf3error}
		\includegraphics[width=0.35\textwidth]{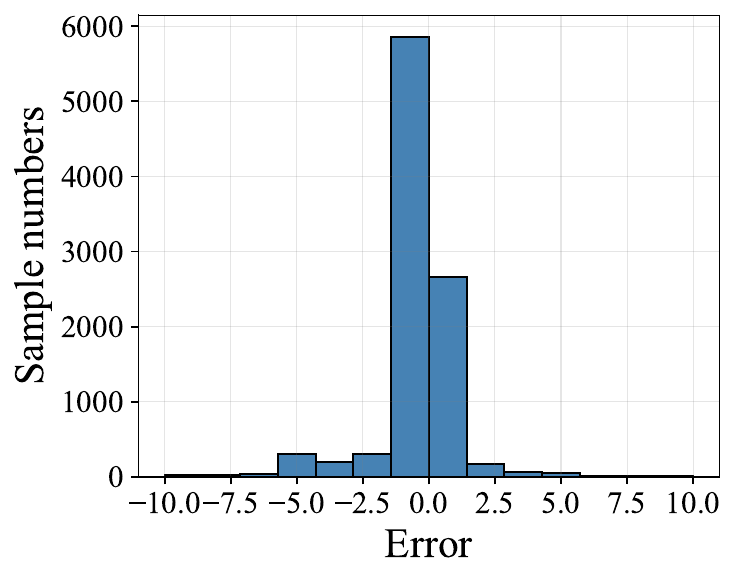}}
	\caption{Error histograms of the DNN in the test set.}
	\label{error}
\end{figure}

The estimation error distributions of ${t_{2,f}}$ and ${t_{3,f}}$ in the test set are shown in Fig. \ref{error}. It can be seen that the estimation error is mainly concentrated and distributed in a range around 0 s, so this DNN can establish a good mapping relationship from the fault state to the time of coasting and the orbiting phases.

\begin{figure}[htbp]
	\centering
	\subfigure[SJTR]{
		\label{fig.SJTR}
		\includegraphics[width=0.35\textwidth]{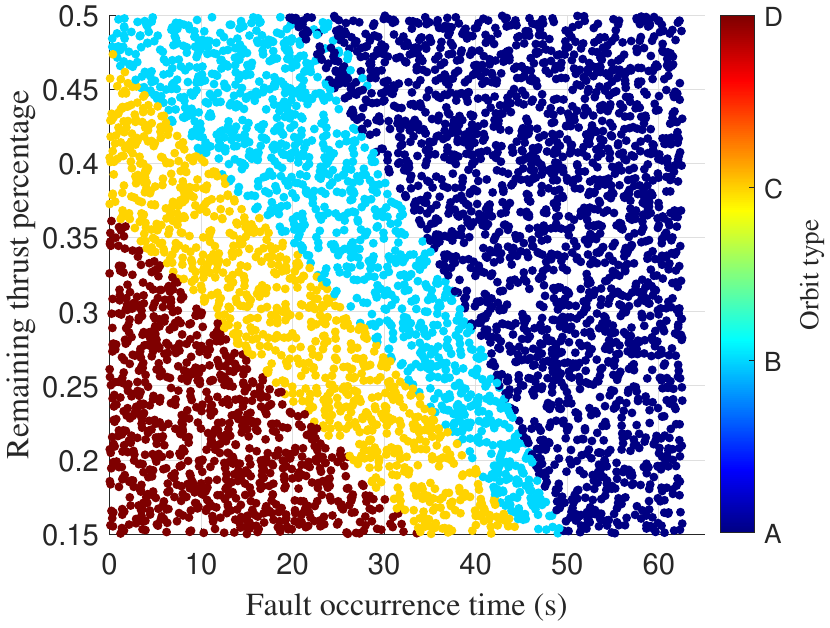}}
	\subfigure[Learning-based warm-start SJTR]{
		\label{fig.SJTR2}
		\includegraphics[width=0.35\textwidth]{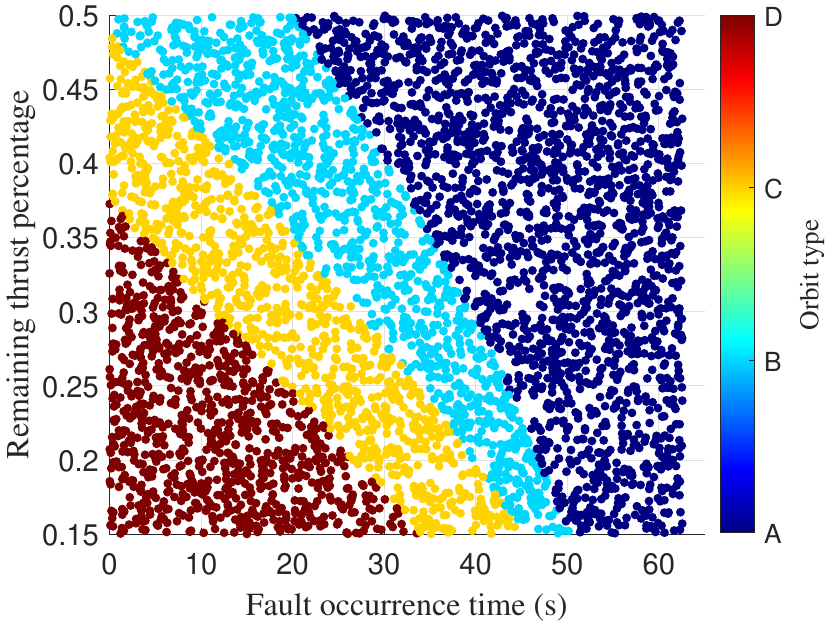}}
	\caption{Orbit type statistical graph.}
	\label{fig.Orbit type}
\end{figure}

To verify the effectiveness of the learning-based warm-start SJTR method, we conducted 5,000 sets of Monte Carlo simulations with the same fault conditions distribution as previously mentioned. When the MAV detects a fault, the initial time guesses are obtained by using the trained network. Subsequently, the SJTR is executed and the corresponding flight trajectory and orbit type are optimized at the same time. The results of the solution using the SJTR method and the learning-based warm-start SJTR method are shown in Fig. \ref{fig.Orbit type}.

In this result, each point represents a fault condition characterized by the combination of fault occurrence time and remaining thrust percentage. The four colors A, B, C, and D correspond to four orbit types: original target orbit, rescue orbit type I, rescue orbit type II, and mission failure, respectively.

The average runtime of the learning-based warm-start SJTR is 0.449 s, and the average number of iterations is 23.357. Compared to the SJTR, the efficiency is improved by 3\%, and the number of iterations is reduced by 1.3. However, the improvement in computational performance is limited because the SJTR already has an excellent convergence performance. As shown in Fig. \ref{fig.SJTR}, the SJTR without using the warm-start approach can result in some unsmooth boundaries and inaccurate orbit type determinations. This is because, in some fault situations corresponding to ambiguous orbit types, poor initial guesses prevent the SJTR from finding a better solution or lead to convergence difficulties. Thus, the learning-based warm-start improves the performance of the algorithm by providing a better initial guess for SJTR.

\subsection{Comparison of Trajectory Replanning Methods}

To demonstrate the effectiveness of the learning-based warm-start SJTR, we compare it with the general trajectory replanning method based on STI through the four different fault cases shown in Table \ref{tab_2}. Since the upgrade and downgrade schemes in the general trajectory replanning method are essentially the same, we only use the downgrade scheme as a comparison with the proposed method. Based on the solution process of the downgrade scheme shown in Fig. \ref{fig.1}, the general trajectory replanning method can be divided into Steps 1-3, corresponding to solving P4, P3, and P2, respectively.
\begin{table}[htbp]
	\centering
	\caption{Cases for 4 orbit types}
	\renewcommand{\arraystretch}{1.2}
	\begin{tabular}{c c c c c c}
		\hline
		\ & Case 1&Case 2& Case 3& Case 4& Unit\\
		\hline
		Orbit type           & A & B & C & D & -\\
		${r}_{x,\text{fail}}$& -288.64  & -283.16  & -300.17  & -295.04  & km\\
		${r}_{y,\text{fail}}$& -3380.76 & -3384.03 & -3374.99 & -3377.26 & km\\
		${r}_{z,\text{fail}}$& 245.32   & 249.15   & 238.75   & 241.28   & km\\
		${v}_{x,\text{fail}}$& 748.27   & 886.07   & 383.60   & 565.85   & m/s\\
		${v}_{y,\text{fail}}$& -432.83  & -545.33  & -137.37  & -284.58  & m/s\\
		${v}_{z,\text{fail}}$& 504.53   & 638.88   & 144.53   & 325.30   & m/s\\
		${m}_{\text{fail}}$  & 255.84   & 234.81   & 320.38   & 286.46   & kg\\
		${t_{\text{fail}}}$  & 30.06    & 36.78    & 9.46     & 20.29    & s\\
		$\eta$               & 46.84    & 24.12    & 34.34    & 20.81    & \%\\
		\hline
	\end{tabular}
	\label{tab_2}
\end{table}

\subsubsection{Case 1} Into the original target orbit (shown in Table \ref{Case 1}).

The result obtained from solving Step 1 indicates that there is still remaining fuel after constraining all terminal conditions. Both methods show that the deviations in altitude $\Delta h_f$, orbital inclination $\Delta i_f$, and longitude of ascending node $\Delta \Omega_f$ from the original target orbit are all zero, which supports the MAV in entering the original target orbit. Based on the optimized final mass of the MAV, it can be seen that compared to the general trajectory replanning method, the proposed method sacrifices some optimality. This characteristic facilitates the joint trajectory replanning method designed in this paper.
\begin{table}[h]
	\centering
	\caption{Comparison of results for case 1}
	\renewcommand{\arraystretch}{1.2}
	\begin{tabular}{c|>{\centering\arraybackslash}p{1.2cm}>{\centering\arraybackslash}p{1.2cm}>{\centering\arraybackslash}p{1.2cm}|c}
		\hline
		\multirow{2}[4]{*}{} & \multicolumn{3}{c|}{General Trajectory Replanning Method} & \multirow{2}[1]{*}{\makecell{Proposed\\Method}} \\
		\cline{2-4}          & Step 1 & Step 2 & Step 3 &  \\
		\hline
		$m_{3,f}$ (kg)    & 75.75 & -     & -     & 75.67 \\
		$\Delta h_f$ (km)  & 0     & -     & -     & 0 \\
		$\Delta i_f$ (deg)  & 0     & -     & -     & 0 \\
		$\Delta \Omega_f$ (deg)  & 0     & -     & -     & 0 \\
		\hline
		Runtime (s)  & \multicolumn{3}{c|}{0.6322}  & 0.5997 \\
		\hline
	\end{tabular}%
	\label{Case 1}%
\end{table}%

\subsubsection{Case 2} Into the rescue orbit type I (shown in Table \ref{Case 2}).

Solving P4 in Step 1 reveals that the fuel required to enter the original target orbit exceeds the actual fuel available. Therefore, Step 2 is initiated, in which the feasible solution obtained by solving P3 belongs to the rescue orbit type I. According to Table \ref{Case 2}, it takes 1.3186 s from detecting the fault occurrence to planning a new trajectory using the general trajectory replanning method, while the proposed method takes only 0.6823 s, and the parameters of the final rescue orbit are not much different.
\begin{table}[h]
	\centering
	\caption{Comparison of results for case 2}
	\renewcommand{\arraystretch}{1.2}
	\begin{tabular}{c|>{\centering\arraybackslash}p{1.2cm}>{\centering\arraybackslash}p{1.2cm}>{\centering\arraybackslash}p{1.2cm}|c}
		\hline
		\multirow{2}[4]{*}{} & \multicolumn{3}{c|}{General Trajectory Replanning Method} & \multirow{2}[1]{*}{\makecell{Proposed\\Method}} \\
		\cline{2-4}          & Step 1 & Step 2 & Step 3 &  \\
		\hline
		$m_{3,f}$ (kg)    & 74.60  & 75.40  & -     & 75.40 \\
		$\Delta h_f$ (km)  & 0     & 0     & -     & 0 \\
		$\Delta i_f$ (deg)  & 0     & -0.49 & -     & -0.498 \\
		$\Delta \Omega_f$ (deg)  & 0     & 1.691 & -     & 1.720 \\
		\hline
		Runtime (s)  & \multicolumn{3}{c|}{1.3186} & 0.6823 \\
		\hline
	\end{tabular}%
	\label{Case 2}%
\end{table}%

\subsubsection{Case 3} Into the rescue orbit type II (shown in Table \ref{Case 3}).

Step 2 of solving this case will not yield a feasible solution because P3 strictly constrains the altitude of the terminal orbit. Therefore, the process moves to Step 3. The final altitude from the target orbit obtained by the general trajectory replanning method and the proposed method are 29.4 km and 31.84 km, respectively, both of which meet the minimum safe orbit requirements. The results correspond to the rescue orbit type II, with the general method producing a better solution. The total time required to solve Steps 1 and 3 is 1.5077 s. Although no feasible solution is found in Step 2, the calculation is still necessary, rendering its runtime statistics meaningless. Thus, solving this case using the general trajectory replanning method requires at least 1.5077 s (denoted as 1.5077+). In contrast, the proposed method only requires 0.6942 s, improving the efficiency of decision-making.

\begin{table}[h]
	\centering
	\caption{Comparison of results for case 3}
	\renewcommand{\arraystretch}{1.2}
	\begin{tabular}{c|>{\centering\arraybackslash}p{1.2cm}>{\centering\arraybackslash}p{1.2cm}>{\centering\arraybackslash}p{1.2cm}|c}
		\hline
		\multirow{2}[4]{*}{} & \multicolumn{3}{c|}{General Trajectory Replanning Method} & \multirow{2}[1]{*}{\makecell{Proposed\\Method}} \\
		\cline{2-4}          & Step 1 & Step 2 & Step 3 &  \\
		\hline
		$m_{3,f}$ (kg)    & 72.22 & Infeasible & 75.40  & 75.40 \\
		$\Delta h_f$ (km)  & 0     &  Infeasible     & -29.40 & -31.84 \\
		$\Delta i_f$ (deg)  & 0     &  Infeasible     & -9.592 & -1.082 \\
		$\Delta \Omega_f$ (deg)  & 0     &  Infeasible     & 0.631 & 4.171 \\
		\hline
		Runtime (s)  & \multicolumn{3}{c|}{1.5077+} & 0.6942 \\
		\hline
	\end{tabular}%
	\label{Case 3}%
\end{table}%

\subsubsection{Case 4} Mission failure (shown in Table \ref{Case 4}).

In this case, the maximum orbital altitude that the MAV can reach, as calculated by solving P2, is less than the safe orbital altitude. As shown in Table \ref{Case 4}, the general trajectory replanning method requires at least 1.4531 s to determine this case, whereas the proposed method only takes 0.6812 s.
\begin{table}[h]
	\centering
	\caption{Comparison of results for case 4}
	\renewcommand{\arraystretch}{1.2}
	\begin{tabular}{c|>{\centering\arraybackslash}p{1.2cm}>{\centering\arraybackslash}p{1.2cm}>{\centering\arraybackslash}p{1.2cm}|c}
		\hline
		\multirow{2}[4]{*}{} & \multicolumn{3}{c|}{General Trajectory Replanning Method} & \multirow{2}[1]{*}{\makecell{Proposed\\Method}} \\
		\cline{2-4}          & Step 1 & Step 2 & Step 3 &  \\
		\hline
		$m_{3,f}$ (kg)    & 68.60 & Infeasible & 75.40  & 75.40 \\
		$\Delta h_f$ (km)  & 0     &   Infeasible    & -70.81 & -75.87 \\
		$\Delta i_f$ (deg)  & 0     &  Infeasible     & -2.101 & -0.950 \\
		$\Delta \Omega_f$ (deg)  & 0     &  Infeasible     & 3.914 & 3.788 \\
		\hline
		Runtime (s)  & \multicolumn{3}{c|}{1.4531+} & 0.6812 \\
		\hline
	\end{tabular}%
	\label{Case 4}%
\end{table}%

In addition to the advantages of computational efficiency, the most important benefit of the proposed method is that it achieves replanning without complex step-by-step judgments. Furthermore, the general trajectory replanning method may encounter the issue of infeasible solution propagation, whereas the proposed method almost never encounters infeasible solutions, which significantly enhances the system's reliability.

\section{CONCLUSIONS}

This paper proposes the SJTR method for optimizing the target orbit and flight trajectory of a MAV after encountering a thrust drop propulsion system fault during its flight. The method integrates the MCPI framework, known for its good convergence performance, and reveals the feature that the optimization problem can adhere to the orbit redecision principle by designing penalty coefficients for terminal constraints. Additionally, by establishing a DNN that maps fault situations to time optimization variables, the SJTR method is equipped with the learning-based warm-start strategy that overcomes the problem of inaccurate orbit type determination. Simulation results show that, compared to the general trajectory replanning method, the proposed method eliminates the need for step-by-step decision-making, offering higher computational efficiency and solution feasibility. It provides a non-optimal but highly reliable solution for hazardous Mars ascent missions.

%

%

\bibliographystyle{IEEEtran}
\bibliography{reference}  

\end{document}